\documentclass[twocolumn,showpacs,preprintnumbers,amsmath,amssymb]{revtex4}
\usepackage{graphicx}
\usepackage{dcolumn}
\usepackage{bm}

\begin{document}
\title{Analysis of $\Lambda_{b}\to$ $p K^{-}$ and $p \pi^{-}$ decays in a flavor changing $Z^{\prime}$ model}
\author{Shuaiwei Wang}
\email{shuaiweiwang@sina.com}
\author{Jinshu Huang}
\author{Genquan Li}
\affiliation{College of Physics $\&$ Electronic Engineering, Nanyang
Normal University, Nanyang 473061, People's Republic of China}

\date{\today}

\begin{abstract}
In this paper, we calculate the branching ratios of $\Lambda_{b}\to
p K^{-}$ and $p \pi^{-}$ decays in the flavor changing  $Z^{\prime}$
model. We find that the branching ratios for these two decay modes
are sensitive to the contributions of $Z^{\prime}$ boson. For
$\Lambda_{b}\to p K^{-}$ decay, if the left-handed couplings are
equal to the right-handed couplings, the branching ratio of this
decay could match up to the currently experimental data for
$\xi_{s}=0.01$ and $-52^{\circ}<\phi^{L}_{s}<132^{\circ}$, or
$\xi_{s}=0.004$ and $0^{\circ}<\phi^{L}_{s}<84^{\circ}$; if only the
left-handed couplings are considered, it could match up to the
experimental data for $\xi_{s}=0.01$ and
$-10^{\circ}<\phi^{L}_{s}<138^{\circ}$. And for $\Lambda_{b}\to p
\pi^{-}$ decay, if the left-handed and right-handed couplings are
equal, the branching ratio of $\Lambda_{b}\to p \pi^{-}$ decay may
be consistent with the currently experimental data with
$\xi_{d}=0.05$ and $-135^{\circ}<\phi^{L}_{d}<43^{\circ}$, if only
the left-handed couplings are considered, it may be consistent with
$\xi_{d}=0.05$ and $-114^{\circ}<\phi^{L}_{d}<8^{\circ}$.
\end{abstract}


\maketitle

\section{\label{sec:level1}Introduction}

Nonleptonic weak decays of heavy baryons arouse people's great interest since they provide a good area to understand the standard model (SM) and search new physics (NP) beyond the SM. For the branching
ratios of $\Lambda_{b}\to p K^{-}$ and $p \pi^{-}$ decays, the CDF Collaboration has presently reported $[1]$
\begin{eqnarray}
Br(\Lambda_{b}\to p K^{-})=(5.6\pm0.8\pm1.5)\times10^{-6},\nonumber \\
Br(\Lambda_{b}\to p \pi^{-})=(3.5\pm0.6\pm0.9)\times10^{-6}.
\end{eqnarray}

In the SM, $\Lambda_{b}\to p K^{-}$ and $\Lambda_{b}\to p \pi^{-}$
decays have been discussed $[2,3]$. The authors used the
generalized and naive factorization approaches to handle the hadronic
matrix elements, respectively. However, the branching ratios for
these two decay modes are significantly lower than the present
experiment values. Specially in Ref. [3], the measured branching
ratio for $\Lambda_{b}\to p K^{-}$ could be easily accommodated in
the fourth quark generation model. In this paper, using the QCD
factorization approach, we shall discuss the branching ratios for
$\Lambda_{b}\to p K^{-}$ and $p \pi^{-}$ decays in a flavor changing
$Z^{\prime}$ model $[4,5]$.

The paper is organized as follows. In the next section we shall discuss the nonleptonic decays of $\Lambda_{b}$ baryon. In Section 3, we shall analyze these decay modes in a flavor changing $Z^{\prime}$ model. Section 4 contains our conclusions.

\section{\label{sec:level2} $\Lambda_{b}\to p K^{-}$ and $p \pi^{-}$ decays}

For the nonleptonic $\Lambda_{b}$ decay modes $\Lambda_{b}\to p K^{-}$ and $p \pi^{-}$, which are induced by the
quark level transition $b\to p\bar{u}u$ ($p=d,s$), the effective Hamiltonian can be written as $[6]$
\begin{equation}
\mathcal{H}_{eff}=\frac{G_F}{\sqrt{2}}\biggl[V_{ub}V_{up}^{*} \sum_{i=1,2}C_{i}(\mu)O_{i}-V_{tb}V_{tp}^{*}
\sum_{i=3}^{10}C_{i}(\mu)O_{i}\biggl],
\end{equation}
where $C_{i}$ are the effective Wilson coefficients and $Q_{i}$ the relevant four-quark operations.

Firstly, we discuss $\Lambda_{b}\to p K^{-}$ decay in the SM. The amplitude for this process in the factorization approximation is
given as $[2]$
\begin{eqnarray}
&& \mathcal{A}(\Lambda_{b}(p)\to p(p^{\prime}) K^{-}(q))=
i\frac{G_F}{\sqrt{2}}f_{K}\bar{u}_{p}(p^{\prime}) \times\nonumber \\&&
\biggl[\biggl(V_{ub}V_{us}^{*}a_{1}
-V_{tb}V_{ts}^{*}(a_{4}+a_{10}+(a_{6}+a_{8})R_{1})\biggl) \nonumber\\
&& \times \biggl(g_{1}(q^{2})(m_{\Lambda_{b}}-m_{p})+
g_{3}(q^{2})m_{K}^{2}\biggl) + \nonumber\\&& \biggl(V_{ub}V_{us}^{*}a_{1}
-V_{tb}V_{ts}^{*}(a_{4}+a_{10}-(a_{6}+a_{8})R_{2})\biggl) \nonumber\\
&& \times \biggl(G_{1}(q^{2})(m_{\Lambda_{b}}+m_{p})-
G_{3}(q^{2})m_{K}^{2}\biggl)\gamma_{5}\biggl]u_{\Lambda_{b}}(p), \nonumber \\
\end{eqnarray}

\vspace{0.2cm}

where
\begin{eqnarray}
&& R_{1}=\frac{2m_{K}^{2}}{(m_{b}-m_{u})(m_{s}+m_{u})},\nonumber \\
&& R_{2}=\frac{2m_{K}^{2}}{(m_{b}+m_{u})(m_{s}+m_{u})}.
\end{eqnarray}

The matrix elements of the various hadronic currents between the initial $\Lambda_{b}$ and final $p$ baryon are parameterized in
terms of various form factors as $[7]$
\begin{eqnarray}
&& <p(p^{\prime})|\bar{s}\gamma^{\mu}b|\Lambda_{b}(p)>=\bar{u}_{p}(p^{\prime})\biggl[g_{1}(q^{2})\gamma_{\mu} \nonumber \\
&& +ig_{2}(q^{2})\sigma^{\mu\nu}q^{\nu}+g_{3}(q^{2})q_{\mu}\biggl]u_{\Lambda_{b}}(p),
\end{eqnarray}
\begin{eqnarray}
&& <p(p^{\prime})|\bar{s}\gamma^{\mu}\gamma_{5}b|\Lambda_{b}(p)>=\bar{u}_{p}(p^{\prime})\biggl[G_{1}(q^{2})\gamma_{\mu} \nonumber \\
 && +iG_{2}(q^{2})\sigma^{\mu\nu}q^{\nu}+G_{3}(q^{2})q_{\mu}\biggl]\gamma_{5}u_{\Lambda_{b}}(p),
\end{eqnarray}
\noindent where $q$ is the momentum transfer, i.e., $q=p-p^{\prime}$. $g_{i}$ and $G_{i}$ are the vector and axial vector form factors, respectively. For the final state of $K^{-}$ meson, it can be written as
\begin{eqnarray}
<K^{-}(q)|\bar{u}\gamma^{\mu}\gamma_{5}u|0>=if_{K}q^{\mu}/\sqrt{2},
\end{eqnarray}
where $f_{K}$ is the decay constant of $K$ meson. The amplitude of $\Lambda_{b}\to p K^{-}$ decay is also written as $[2]$
\begin{eqnarray}
\mathcal{A}(\Lambda_{b}(p)\to p(p^{\prime}) K^{-}(q))=i \bar{u}_{p}(p^{\prime})(A+B\gamma_{5})u_{\Lambda_{b}}(p),
\end{eqnarray}
with
\begin{widetext}
\begin{eqnarray}
A&=&\frac{G_F}{\sqrt{2}}f_{K}\biggl(V_{ub}V_{us}^{*}a_{1}
-V_{tb}V_{ts}^{*}(a_{4}+a_{10}+(a_{6}+a_{8})R_{1})\biggl) \biggl(g_{1}(q^{2})(m_{\Lambda_{b}}-m_{p})+
g_{3}(q^{2})m_{K}^{2}\biggl),\nonumber \\
B&=&\frac{G_F}{\sqrt{2}}f_{K}\biggl(V_{ub}V_{us}^{*}a_{1}
-V_{tb}V_{ts}^{*}(a_{4}+a_{10}-(a_{6}+ a_{8})R_{2})\biggl) \biggl(G_{1}(q^{2})(m_{\Lambda_{b}}+m_{p})-
G_{3}(q^{2})m_{K}^{2}\biggl),
\end{eqnarray}
thus the branching ratio for this process can be written as $[8]$
\begin{eqnarray}
Br(\Lambda_{b}\to p K^{-})&=& \tau_{\Lambda_{b}}\frac{p_{cm}}{8\pi}\biggl[\frac{(m_{\Lambda_{b}}+m_{p})^2
-m_{K}^{2}}{m_{\Lambda_{b}}^2}|A|^{2}+ \frac{(m_{\Lambda_{b}}-m_{p})^2
-m_{K}^{2}}{m_{\Lambda_{b}}^2}|B|^{2}\biggl].
\end{eqnarray}

Similarly, the amplitude of $\Lambda_{b}\to p \pi^{-}$ decay may be given as $[6]$
\begin{eqnarray}
&& \mathcal{A}(\Lambda_{b}\to p
\pi^{-})= i\frac{G_F}{\sqrt{2}}f_{\pi}\bar{u}_{p}(p^{\prime})\biggl[\biggl(V_{ub}V_{ud}^{*}a_{1}
- V_{tb}V_{td}^{*}(a_{4}+a_{10}+(a_{6}+a_{8})R_{1})\biggl) \biggl(g_{1}(q^{2})(m_{\Lambda_{b}}-m_{p})+
g_{3}  \nonumber \\
&&  \times (q^{2})m_{\pi}^{2}\biggl) + \biggl(V_{ub}V_{ud}^{*}a_{1}
-V_{tb}V_{td}^{*}(a_{4}+a_{10}-(a_{6}+a_{8})R_{2})\biggl) \biggl(G_{1}(q^{2})(m_{\Lambda_{b}}+m_{p})-
G_{3}(q^{2})m_{\pi}^{2}\biggl)\gamma_{5}\biggl]u_{\Lambda_{b}}(p),
\end{eqnarray}
\end{widetext}

\noindent where
\begin{eqnarray}
R_{1}=\frac{2m_{\pi}^{2}}{(m_{b}-m_{u})(m_{d}+m_{u})}, \nonumber \\ R_{2}=\frac{2m_{\pi}^{2}}{(m_{b}+m_{u})(m_{d}+m_{u})},
\end{eqnarray}
and we can also obtain the branching ratio for $\Lambda_{b}\to p
\pi^{-}$ decay using Eq. (10).

For numerical analysis, the form factors in the space-like region can be parameterized by the following three-parameters fit [9]
\begin{equation}
F_{i}(q^{2})=\frac{F_{i}(0)}{(1-q^{2}/m^{2}_{\Lambda_{b}})(1-a(q^{2}/m^{2}_{\Lambda_{b}})+b(q^{4}/m^{4}_{\Lambda_{b}}))},
\end{equation}
\noindent where the values of the parameters $F_{i}(0)$, $a$ and $b$ have been presented in Table 1. We have already used these parameters to handle the $\Lambda_{b}\to \Lambda \rho^{0}$ , $p K^{*-}$ and $p \rho^{-}$ decays in Ref. [10]. The masses of these  particles, the decay constants of $\pi$, $K$ mesons, the lifetime of $\Lambda_{b}$ baryon, and the CKM elements can be found in  Refs. [11] and [12].

\vspace{0.2cm}

\begin{center}
{\small Table 1. The form factors of $\Lambda_{b}\to p$ transition.}

\vspace{0.1cm}

\begin{tabular}{c|ccc}
\hline
$F$&$F(0)$& $a$ & $b$ \\
\hline
$g_{1}$&0.1131&1.70&1.60\\
$g_{3}$&-0.0356&2.5&2.57\\
$G_{1}$&0.1112&1.65&1.60\\
$G_{3}$&-0.0097&2.8&2.7\\
\hline
\end{tabular}
\end{center}

\vspace{0.2cm}

Combining the above formulas, we can obtain the branching ratios of $\Lambda_{b}\to p K^{-}$ and $p \pi^{-}$ decays in the SM
\begin{eqnarray}
Br(\Lambda_{b}\to p K^{-})&=& (2.73^{+0.17}_{-0.10})\times10^{-6},\nonumber \\
Br(\Lambda_{b}\to p \pi^{-})&=& (1.36^{+0.14}_{-0.09})\times10^{-6}.
\end{eqnarray}

Comparing with Eq. (1), we can find that the branching ratios of $\Lambda_{b} \to p K^{-}$ and $p \pi^{-}$ decays are lower than the present experimental values at about $4\sigma$ level.

\section{\label{sec:level3} Numerical analysis in the flavor changing $Z^{\prime}$ model}

A flavor changing $Z^{\prime}$ model can lead to FCNC processes at the tree level due to the non-diagonal chiral coupling matrix, and the formalism of the model has been discussed explicitly in Refs. [4, 5]. Here we shall briefly review the ingredients needed in this paper.

If we neglect the right-handed flavor changing couplings, the effective Hamiltonian of the $b\to p\bar{u}u$ transition mediated by the $Z^{\prime}$ can be written as $[13]$
\begin{equation}
H_{eff}^{Z^{\prime}}=\frac{4G_{F}}{\sqrt{2}} (\frac{g^{\prime}M_{Z}}{gM_{Z^{\prime}}})^{2}
B_{pb}^{L}(B_{uu}^{L}O_{9}+B_{uu}^{R}O_{7})+ \ {\rm h.c.},
\end{equation}
where $p=d,s$ and $M_{Z^{\prime}}$ denotes the mass of new $Z^{\prime}$ gauge boson. Thus the additional contributions to the SM Wilson coefficients at the
$M_{W}$ scale are
\begin{equation}
\triangle C_{9,7}=\frac{4}{V_{tb}V_{tp}^{*}}(\frac{g^{\prime}M_{Z}}{gM_{Z^{\prime}}})^{2}
B_{pb}^{L}B_{uu}^{L,R},
\end{equation}
where our assumption that there is no significant renormalization group (RG) running effect between $M_{Z^{\prime}}$ and $M_{W}$ scales, and the RG evolution of the modified wilson coefficients is exactly the same as the ones in the SM $[14]$.

The diagonal elements of the effective matrices $B_{uu}^{L,R}$ are real due to the Hermiticity of the effective Hamiltonian. However, the off-diagonal ones of $B_{pb}^{L}$ may contain a new weak phase $\phi^{L}_{p}$. Moreover, Eq. (16) can also be simplified as
\begin{equation}
\triangle
C_{9,7}=4\frac{|V_{tb}V_{tp}^{*}|}{V_{tb}V_{tp}^{*}}\xi_{p}^{LL,LR}e^{i\phi^{L}_{p}},
\end{equation}
where the real NP parameters $\xi_{p}^{LL,LR}$ and $\phi_{p}^{L}$
are defined respectively as
\begin{eqnarray}
\xi_{p}^{LL,LR}&=&(\frac{g^{\prime}M_{Z}}{gM_{Z^{\prime}}})^{2}|
\frac{B_{pb}^{L}B_{uu}^{L,R}}{V_{tb}V_{tp}^{*}}|,\nonumber\\
\phi_{p}^{L}&=& Arg{[B_{pb}^{L}]}.
\end{eqnarray}

Obviously, the total decay amplitudes of $\Lambda_{b}\to p K^{-}$ and $p \pi^{-}$ decays depend on three additional real parameters $\xi_{p}^{LL}$, $\xi_{p}^{LR}$ and $\phi_{p}^{L}$ when the contributions coming from the $Z^{\prime}$
boson are considered.

For the sake of simplicity, we shall take the following two cases to estimate the effects of new $Z^{\prime}$ gauge boson for $\Lambda_{b} \to p K^{-}$ decay. Case I: considering the left-handed and right-handed couplings are equal, i.e., $\xi_{s}^{LL}=\xi_{s}^{LR}=\xi_{s}$; and Case II: only considering
the left-handed couplings and neglecting the right-handed couplings, i.e., $\xi_{s}^{LL}=\xi_{s},\ \xi_{s}^{LR}=0$. For each case, We shall take $\xi_{s}=0.01$ and $0.004$ as representative values for numerical analysis as in Ref. [15].

\begin{figure}
\includegraphics{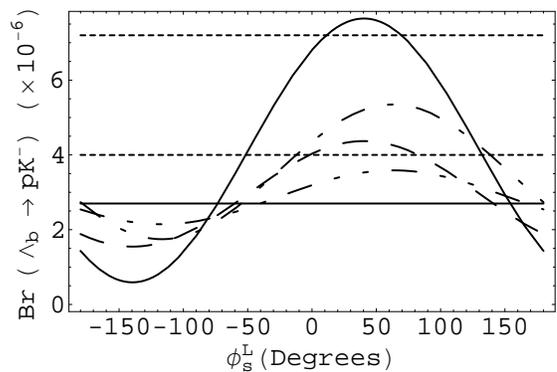}
\caption{\label{fig:eps1} he branch ratio $Br(\Lambda_{b}\to p K^{-})$ versus the weak phase $\phi^{L}_{s}$. The horizontal lines are the branching ratios in the SM, the dotted curves are the currently experimental data of $Br(\Lambda_{b}\to p K^{-})$ within $2 \sigma$. The solid and dashed curves for Case I with $\xi_{s}^{LL}=\xi_{s}^{LR}=0.01,\ 0.004$, and the single-dot-dashed and double-dot-dashed curves for Case II with $\xi_{s}^{LL}=0.01,\ 0.004$, respectively. }
\end{figure}

We plot the branch ratio $Br(\Lambda_{b}\to p K^{-})$ versus the NP weak phase $\phi_{s}^{L}$ given in Fig. 1. From this figure, we can find that (i) for Case I, we obtain $-52^{\circ}<\phi^{L}_{s}<132^{\circ}$ and $0^{\circ}<\phi^{L}_{s}<84^{\circ}$ for $\xi_{s}=0.01$ and $0.004$, respectively; (ii) for Case II, we can get $-10^{\circ}<\phi^{L}_{s}<138^{\circ}$ for $\xi_{s}=0.01$, however, when we take $\xi_{s}^{LL}=\xi_{s}=0.004$, the branching ratio for $\Lambda_{b}\to p K^{-}$ decay could not be consistent with the currently experimental data no matter what values $\phi_{s}^{L}$ may take. Therefore, the branching ratio of $\Lambda_{b}\to p K^{-}$ decay may match up to the currently experimental data for Case I with $\xi_{s}=0.01$ and $0.004$ and for Case II with $\xi_{s}=0.01$.

\begin{figure}
\includegraphics{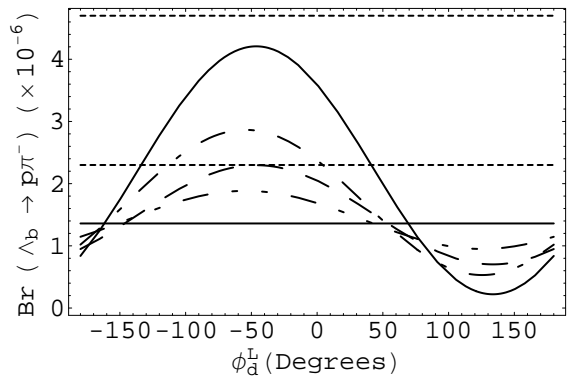}
\caption{\label{fig:eps2} The curve of $Br(\Lambda_{b}\to p \pi^{-})$ vs. the weak phase $\phi^{L}_{d}$. The horizontal lines and  the dotted curves are the same as in Fig. 1 but for $Br(\Lambda_{b}\to p \pi^{-})$. The solid and dashed curves for Case I with $\xi_{d}^{LL}=\xi_{d}^{LR}=0.05,\ 0.02$, and the single-dot-dashed and double-dot-dashed curves for Case II with $\xi_{d}^{LL}=0.05,\ 0.02$, respectively.}
\end{figure}

For $\Lambda_{b}\to p \pi^{-}$ decay, we also take the two cases, Case I: $\xi_{d}^{LL}=\xi_{d}^{LR}=\xi_{d}$ and Case II:
$\xi_{d}^{LL}=\xi_{d},\ \xi_{d}^{LR}=0$. We shall take $\xi_{d}=0.05$ and $0.02$ as representative values for numerical analysis according to the relation $\xi_{d}^{LL,LR}\simeq 5 \xi_{s}^{LL,LR}$ obtained in Ref. [16]. We plot the $Br(\Lambda_{b}\to p\pi^{-})$ versus the NP weak phase $\phi_{d}^{L}$ in Fig. 2. From this figure, it is found that, (i) when $\xi_{d}=0.05$, we can obtain $-135^{\circ}<\phi^{L}_{d}<43^{\circ}$ for Case I and $-114^{\circ}<\phi^{L}_{d}<8^{\circ}$ for Case II, respectively; so the branching ratio of $\Lambda_{b}\to p \pi^{-}$ decay may be consistent with the currently experimental data when $\phi_{d}^{L}$ take values from these regions, (ii) when $\xi_{d}=0.02$, whether the right-handed couplings are considered or not, the branching ratio for $\Lambda_{b} \to p \pi^{-}$ decay could not be enhanced to be consistent with the currently experimental data no matter what values $\phi_{d}^{L}$ may take.

\section{\label{sec:level4} Conclusions}

In this paper we have investigated the effects of $\Lambda_{b}\to p K^{-}$ and $p \pi^{-}$ decays in the flavor changing $Z^{\prime}$ model. For $\Lambda_{b}\to p K^{-}$ decay, We find that if the left-handed couplings are equal to the right-handed couplings, the branching ratio of $\Lambda_{b}\to p K^{-}$ decay could match up to the currently experimental data for $\xi_{s}=0.01$ and $-52^{\circ}<\phi^{L}_{s}<132^{\circ}$, or $\xi_{s}=0.004$ and $0^{\circ}<\phi^{L}_{s}<84^{\circ}$; if only the left-handed couplings are considered, it could match up to the experimental data for $\xi_{s}=0.01$ and $-10^{\circ}<\phi^{L}_{s}<138^{\circ}$, but when
$\xi_{s}=0.004$, it will not be consistent with the currently experimental data no matter what values $\phi_{s}^{L}$ may take. And for $\Lambda_{b}\to p \pi^{-}$ decay, our conclusions are that, if the left-handed and right-handed couplings are equal, the branching ratio of $\Lambda_{b}\to p \pi^{-}$ decay may be consistent with the currently experimental data with $\xi_{d}=0.05$ and $-135^{\circ}<\phi^{L}_{d}<43^{\circ}$, if only the left-handed couplings are considered, it may be consistent with the currently experimental data with $\xi_{d}=0.05$ and $-114^{\circ}<\phi^{L}_{d}<8^{\circ}$, but when $\xi_{d}=0.02$, whether the right-handed couplings are considered or not, the branching ratio for $\Lambda_{b} \to p \pi^{-}$ decay could not be enhanced to be consistent with the currently experimental data no matter what values $\phi_{d}^{L}$ may take.

\section*{ACKNOWLEDGMENTS}

This project was supported by National Natural Science Foundation of China under Grant Nos. 10947020, 10975047 and 10979008, Natural Science Foundation
of He'nan Province under Grant No. 102300410210, Program for Science and Technology Innovation Talents in Universities of He'nan
Province under Grant No. 2012HATIT030, and Youth Foundation of Nanyang Normal University under Grant No. QN2011010.

\end{document}